\documentclass[final,times,5p,xtwocolumn]{elsarticle}
\usepackage[latin9]{inputenc}
\usepackage{float}
\usepackage{amsmath}
\usepackage{amssymb}
\usepackage{graphicx}
\usepackage{esint}
\usepackage[unicode=true,bookmarks=false,breaklinks=false,
pdfborder={0 0 1},backref=section,colorlinks=false]{hyperref}
\makeatletter
\usepackage{latexsym}
\usepackage{epsfig}
\usepackage{amsfonts}
\usepackage{dcolumn}
\usepackage{bm}

\def\br{\begin{eqnarray}}
\def\er{\end{eqnarray}}
\def\be{\begin{equation}}
\def\ee{\end{equation}}

\journal{Physics Letters B}

\usepackage{flushend}

\makeatother

\begin{document}

\title{QCD chiral symmetry restoration with a large number of quarks in
a model with a confining propagator and dynamically massive gluons}

\author[a1,a2]{R. M. Capdevilla}

\author[a3]{A. Doff}

\author[a2,a4]{A. A. Natale}

\address[a1]{Physics Department, University of Notre Dame, 225 Nieuwland Science
Hall, Notre Dame, South Bend, IN 46556, United States,}

\address[a2]{Instituto de Física Teórica, UNESP - Universidade Estadual Paulista,
Rua Dr. Bento T. Ferraz, 271, Bloco II, 01140-070 São Paulo, SP, Brasil}

\address[a3]{Universidade Tecnológica Federal do Paraná - UTFPR - DAFIS, Av.
Monteiro Lobato Km 04, 84016-210 Ponta Grossa, PR, Brazil}

\address[a4]{Centro de Ciências Naturais e Humanas, Universidade Federal do ABC,
09210-170 Santo André - SP, Brasil}

\begin{abstract}
Considering a QCD chiral symmetry breaking model where the gap equation
contains an effective confining propagator and a dressed gluon propagator
with a dynamically generated mass, we verify that the chiral symmetry
is restored for a large number of quarks $n_{f}\approx 7-13$. We
discuss the uncertainty in the results, that is related to the determination
of the string tension ($K_{F}$), appearing in the confining propagator,
and the effective gluon mass ($m_{g}$) at large $n_{f}$.
 \end{abstract}

\begin{keyword}
Chiral symmetry restoration, effective gluon mass, confinement. 
\end{keyword}

\maketitle

Chiral symmetry breaking (csb) is one of the main QCD characteristics.
It is well understood from the phenomenological point of view; where
a quark condensate $\left\langle \bar{q}q\right\rangle $ has a vacuum
expectation value at the scale of few hundred MeV, and a dynamical
quark mass is generated at the same scale. Therefore, the chiral symmetry
is broken and the (almost)massless (pseudo)Goldstone pions are generated.
The knowledge of the csb mechanism in strongly interacting gauge theories
will be complete when we know how this symmetry is broken or restored
as we vary the number of fermions, the fermionic representations,
the temperature or the chemical potential, and how this symmetry is
related to confinement.

In the last years lattice simulations have contributed considerably
to the understanding of a possible connection between csb and confinement,
where center vortices play a fundamental role. In the $SU(2)$ case
the artificial center vortices removal also implies a recovery of
the chiral symmetry \cite{lat1,lat2,lat3}. The relation between
vortices and csb is discussed at length in Ref. \cite{fa}, and follows
old proposals that confinement and csb are intimately connected \cite{ca,ba,co}.
There are also lattice results indicating no direct one-to-one correspondence
between confinement and csb in QCD \cite{su}. Although these results
appear to be in conflict, they may not be fully contradictory according
to the ideas first delineated by Cornwall in Ref. \cite{co1}, where
it was observed that confinement is necessary for csb when quarks
are in the fundamental representation, but the symmetry breaking happens
through one-dressed-gluon exchange if we consider quarks in the adjoint
representation. This means that we may have competing mechanisms as
proposed in the model of Ref. \cite{co2}, which was further studied
in Refs. \cite{us1,us2,us3}.

The csb model of Ref. \cite{co2} describe a gap equation that contains
an effective confining propagator and a dressed gluon propagator with
a dynamically generated mass. To grasp the idea behind the model we
can first discuss what happens with the one-dressed massive gluon
exchange. The effects of dynamical gluon mass generation have been
discussed in Ref. \cite{co1}, and implies a frozen coupling constant
given by 

\begin{equation}
\bar{g}^{2}(k^{2})=\frac{1}{b\ln[(k^{2}+4m_{g}^{2})/\Lambda^{2}]},
\label{eq: eq1}
\end{equation}

where $b=(11N-4n_{f}T(R))/48\pi^{2}$ for the $SU(N)$ group with
$n_{f}$ flavors, and $T(R)$ is connected to the quadratic Casimir
eigenvalue $C_{2}(R)$ for fermions in one specific representation
($R$) of the gauge group. The infrared value of this coupling obtained
through the functional Schrödinger equation \cite{co1} and through
phenomenological analysis \cite{nat} is small. This fact together
with a dynamically massive gluon propagator that roughly behaves as
$1/[k^{2}+m_{g}^{2}]$ in the infrared (IR), cause a damping in the
gap equation erasing the possibility to generate a phenomenologically
acceptable dynamical mass for quarks in the fundamental representation
\cite{co1,ha,hab}. These results have been obtained based on the
Schwinger-Dyson calculations for the gluon propagator \cite{co3,papa,papa1,papa2},
which are in agreement with lattice simulations for the gluon propagator
\cite{cuc}. Note that the study of the gap equation with quarks
in the adjoint representation does lead to csb due to the larger Casimir
operator appearing in the gluon exchange \cite{co1,us1}.

The difficulties to produce csb at the desired level, as discussed
above, led to new approaches to study csb in the context of Schwinger-Dyson
equations (SDE). One of them considers the one-dressed gluon exchange
making use of a gluon propagator described by the lattice data, which
is less damped at intermediate momenta than the one obtained with
the SDE, and with a larger value for the quark-gluon vertex due to
a possible enhancement of the quark-ghost scattering kernel \cite{papa3}.
The other is the model of Ref. \cite{co2} that we addressed before,
where csb for quarks in the fundamental representation is essentially
triggered by a confining propagator, and the main purpose of this
work is to determine how the chiral symmetry is recovered in this
model when we increase the number of fermions.

The confining propagator used in Refs. \cite{co2,us1,us2,us3} is
giving by 

\begin{equation}
D_{eff}^{\mu\nu}(k)\equiv\delta^{\mu\nu}D_{eff}(k)\,;\qquad D_{eff}(k)=\frac{8\pi K_{F}}{(k^{2}+m^{2})^{2}},
\label{eq: eq2}
\end{equation}

where $m$, which is related with the dynamical quark mass $M$, not
only cures the IR singularities of the $1/k^{4}$ propagator, but
also contributes with a negative term to the effective Hamiltonian,
which is crucial to generate the massless pions associated to the
csb. This entropic quality of this propagator has been stressed in
Ref. \cite{cornwall4}.

In QCD we expect that quarks interact through a linear potential proportional
to the string tension $K_{F}$, at least up to a certain distance
\cite{bali}. This behavior is not observed if we perform the Fourier
transform of the time-time component of the dynamically massive gluon
propagator obtained in lattice simulations or in SDE solutions, however
this is a property of Eq.(\ref{eq: eq2}). As far as we know there
is no evidence that the dynamically massive gluon propagator may generate
such linear potential.

A confinement scenario, fully described in \cite{co3}, claims that
dynamical mass generation in QCD lead to an effective theory where
the gluons acquire an effective mass, and consequently this theory
has vortex solutions which are responsible for confinement. This scenario
is consistent with the lattice simulations, where center vortices
seem to be necessary for csb \cite{lat1,lat2,lat3}. In this way
vortices appear in the effective dynamically massive theory, and not
in the QCD Lagrangian and consequently in the SDE. This is the main
reason for the introduction, by hand, of the effective propagator
shown in Eq.(\ref{eq: eq2}) into the gap equation.

The complete gap equation that we shall consider is 

\begin{multline}
M\left(p^{2}\right)=\int\frac{d^{4}k}{(2\pi)^{4}}\left\{ \frac{32\pi K_{F}}{\left[(p-k)^{2}+m^{2}\right]^{2}}\right.+\\
+\left.\frac{3C_{2}\bar{g}^{2}\left(p-k\right)}{(p-k)^{2}+m_{g}^{2}\left(k^{2}\right)}\right\} \frac{M\left(k^{2}\right)}{k^{2}+M^{2}\left(k^{2}\right)},
\label{eq: eq3}
\end{multline}

where we consider a simple fit for the running gluon mass discussed
in \citep{aguina} 

\begin{equation}
m_{g}^{2}\left(k^{2}\right)=\frac{m_{g}^{4}}{k^{2}+m_{g}^{2}}.
\label{eq: eq4}
\end{equation}

It can be proven that the entropic parameter $(m)$ of Eqs.(\ref{eq: eq2})
and (\ref{eq: eq3})has to be proportional to the string tension.
For instance, in \cite{co2} the condition to generate massless pions
is $M^{2}=9K_{F}/4\pi^{2}$, and it was also verified that naturally
$m\approx M$. In Ref. \cite{us1} the bifurcation condition for
the complete gap equation was studied and we can assume 

\begin{equation}
m^{2}=\kappa K_{F},
\label{eq: eq5}
\end{equation}

where $\kappa\approx0.18$ imply in reasonable values of the dynamical
quark mass ($200{\rm MeV}<M<300{\rm MeV}$) for $n_{f}=2$. With these
values it was shown in Refs. \cite{co2} and \cite{us1} that we can describe
the light quark phenomenological parameters, for example, the values
of the quark condensate and the pion decay constant.

There is another important reason to consider the complete gap equation
model of Eq.(\ref{eq: eq3}) with a confining propagator and one-dressed-gluon
exchange. We have observed that for quarks in the fundamental representation
approximately $95\%$ of the dynamical quarks mass is generated by
the confining propagator and $5\%$ by the dynamically massive gluon
exchange, while for quarks in the adjoint representation the result
is exactly the opposite \cite{us1,us3}. This is what we meant in
the beginning by competing mechanisms, and may explain the apparently
contradictory lattice results quoted in the second paragraph, where
csb is related to confinement due to vortices, but no one-to-one relation
between csb and confinement is found in other simulations.

The model that we discuss in this work has also a clear difference
with the scenario in which all the symmetry breaking is generated
by gluon exchange, i.e. when the gap equation has only the dressed-gluon
exchange and a more sophisticated vertex function. If we just assume
naive Casimir scaling for the interaction in the SDE, it is clear
that csb for quarks in different representations will be increased
accordingly to the value of this operator, which will appear in all
Green's functions implicit in the gap equation, and this fact may
be tested by lattice simulations. For instance, if the csb of quarks
in the fundamental representation seems to be enhanced in the one-dressed
gluon exchange approach, the csb for quarks in the adjoint representation
will be enhanced even more. To further strength the qualities of the
model of Ref. \cite{co2}, we comment in the sequence how it can
explain the difference between the chiral transition of fundamental
and adjoint quarks \cite{us3}, without appealing for an enhancement
of the quark-gluon vertex and without the use of the lattice gluon
propagator.

It is known that the chiral symmetry restoration at finite temperature
in QCD with two quark flavors in the fundamental representation is
connected to the de-confinement transition \cite{bazavov,aoki}. Whereas,
when quarks are in the adjoint representation, it has been found that
the chiral transition happens at a temperature ($T_{c}$) higher than
the de-confinement temperature ($T_{d}$) \cite{karsch,engels,bilgici},
where the ratio between these temperatures for adjoint quarks is giving
by \cite{karsch}: ${T_{c}}/{T_{d}}\approx7.7\pm2.1$, and factors
of the same order were found in \cite{engels,bilgici}. Working with
Eq.(\ref{eq: eq3}) we have been able to explain this difference in
\cite{us3}, which is basically related to the different contributions
of the two propagators present in the gap equation. In the realm of
SDE we are not aware of other explanation of this difference in the
chiral transition, what may be a signal of the model success. The
main uncertainty in the calculation of Ref. \cite{us3} is the poor
knowledge of the effective gluon mass for quarks in the adjoint representation.

The chiral symmetry restoration does not happen exclusively with the
increase of the temperature. Recent lattice calculations are indicating
that the chiral symmetry is also restored with the increase of the
number of flavors \cite{for,tom}. In a gap calculation considering
only one-gluon dynamically massive exchange it was observed that the
chiral symmetry is recovered when the number of quarks is $n_{q}\approx8$
\cite{bas}, and our intention is to verify this effect in the model
of Ref. \cite{co2}. It is important to mention that our calculation
makes use of the rainbow approximation, a simple fit of the phenomenological
running of the dynamical gluon mass, the confining effective propagator
with an entropic parameter that follows Eq.(\ref{eq: eq5}), and,
of course, we work in the Landau gauge. The main problem to calculate
how the solution of Eq.(\ref{eq: eq3}) varies with $n_{f}$ (or $n_{q}$),
as stated in the previous paragraph in the finite temperature case,
is the poor knowledge that we have about the variation of $K_{F}$
and $m_{g}$ with the number of quarks.

\begin{figure}[h]
\includegraphics[scale=0.65]{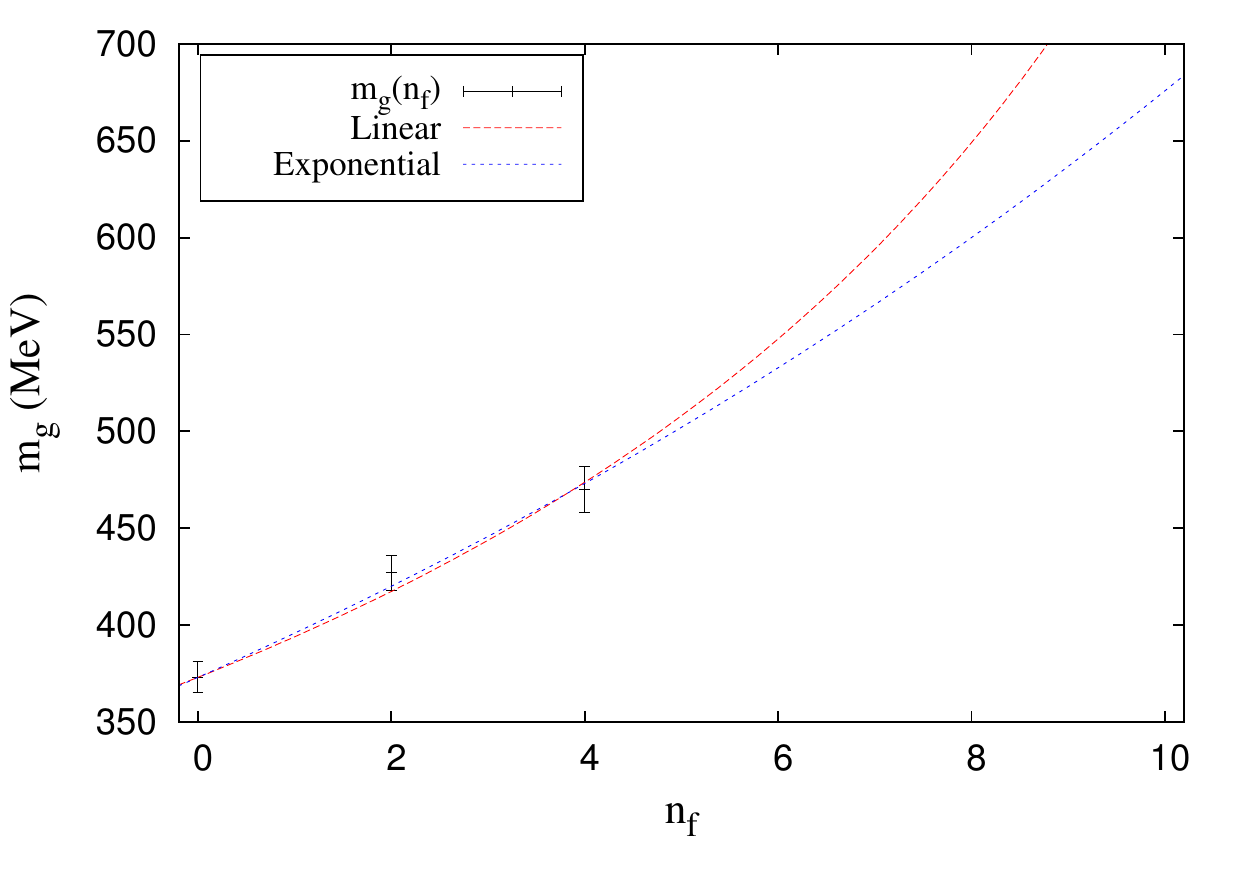} 
\caption{Dynamical gluon mass infrared value as a function of the number of
flavors. The mass is extrapolated for large $n_{f}$ values according
the fits of Eq.(\ref{eq: eq6}).}
\label{fig1} 
\end{figure}

In order to study how the solutions of Eq.(\ref{eq: eq3}) vary as
we change the number of flavors ($n_{f}$) we need to know what happens
with $K_{F}$ and $m_{g}$ as $n_{f}$ is increased. The infrared
value of the dynamical gluon mass was recently determined in lattice
simulations for a small number of flavors \cite{bas}, and the variation
of the string tension with $n_{f}$, also for a small number of quarks,
was discussed in Ref. \cite{kar}. Therefore it will be necessary
to extrapolate the known $K_{F}$ and $m_{g}$ values for large $n_{f}$.
This will be the main uncertainty introduced in the results that we
shall present. According to Ref. \cite{bas} the values for the infrared
dynamical gluon mass are $373(8)$, $427(9)$, $470(12)$ MeV respectively
for $n_{f}=0$, $2$, $4$ quarks. We plot in Fig.(\ref{fig1}) the
linear and exponential curves used to extrapolate these values, whose
best fits are given by 

\begin{equation}
\begin{array}{c}
m_{g}(n_{f})=\frac{0.5}{\left(1-0.053\, n_{f}\right)}{\rm GeV},\\
\\
m_{g}(n_{f})=0.5e^{0.059\, n_{f}}{\rm GeV}.
\end{array}
\label{eq: eq6}
\end{equation}

Note that the fits were normalized in order to match the phenomenologically
acceptable value $m_{g}\approx2\Lambda_{QCD}\approx500{\rm MeV}$
that has been obtained in Refs. \cite{nat,co3,luna}.

To know how the string tension ($K_{F}$) for quarks in the fundamental
representation varies with the number of quarks we followed Ref. \cite{kar}
and assumed $K_{F}=0.18(1)$, $0.17(2)$, $0.14(1)$, $0.12(3)$ GeV$^{2}$
respectively for $n_{f}=0$, $2$, $3$, $4$ flavors. We them performed
two different fits (in units of ${\rm GeV^{2}}$): 

\begin{equation}
\begin{array}{c}
{\rm Gaussian:\qquad}K_{F}(n_{f})=0.183\,{\rm exp}\left[-\left(\frac{n_{f}}{6.1}\right)^{2}\right],\\
\\
{\rm Linear:}\qquad K_{F}(n_{f})=0.187-0.015n_{f},\\
\\
{\rm Power-1:}\qquad K_{F}(n_{f})=0.029(10.21-n_{f})^{0.8},\\
\\
{\rm Power-2:}\qquad K_{F}(n_{f})=0.07(7.0-n_{f})^{0.5}.
\end{array}
\label{fits}
\end{equation}

\begin{figure}[h]
\includegraphics[scale=0.65]{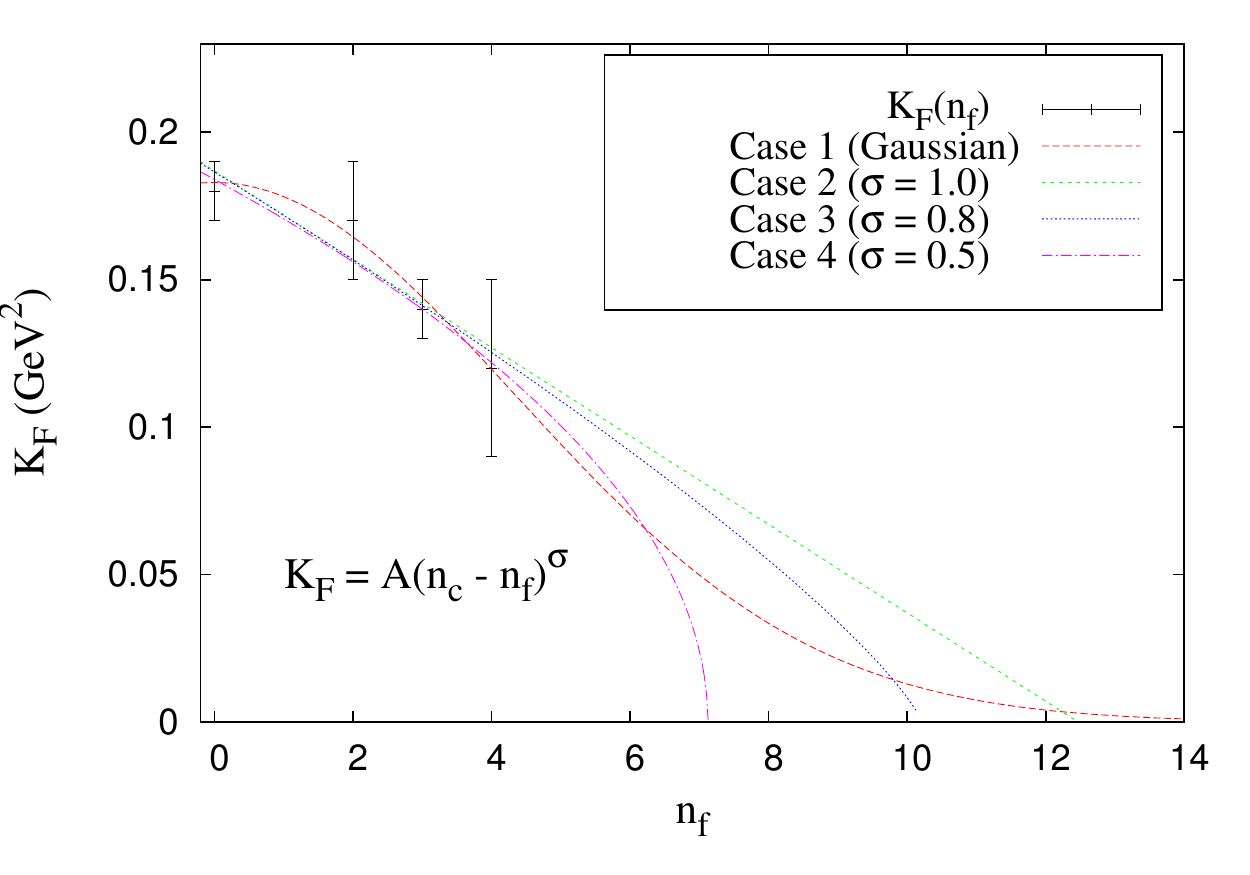} 
\caption{Fits, giving by Eq.(\ref{fits}), of the string tension as a function
of the number of flavors.}
\label{fig2} 
\end{figure}

These fits, hereafter termed as Cases $1$, to $4$, are shown
in Fig.(\ref{fig2}), where we may see
that, for instance in the case 3, the string tension goes to zero as $n_{f}\approx10$. 

\begin{figure}[h]
\includegraphics[scale=0.65]{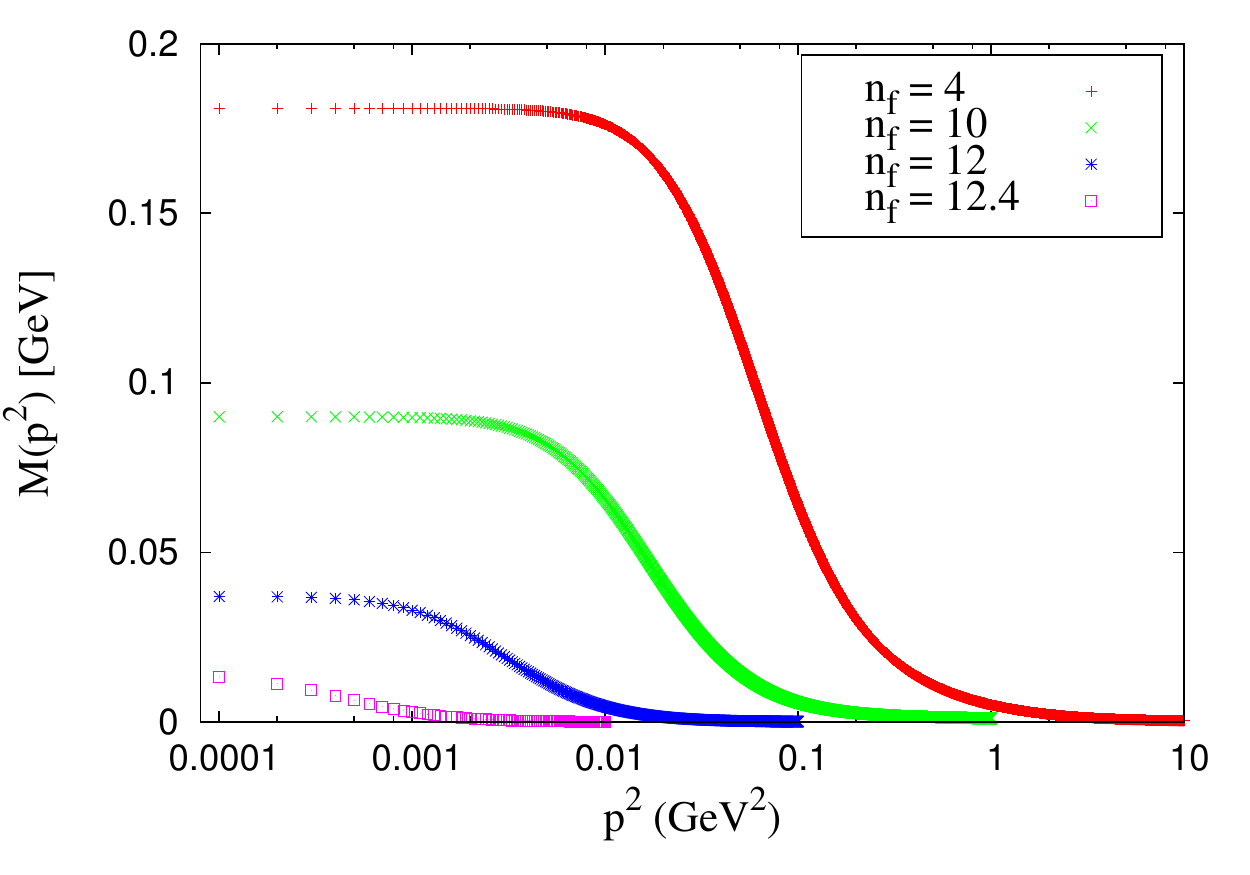} 
\caption{Dynamical quark masses as a function of the momenta for different
number of flavors, computed in the Case 2 (Linear fit for the string
tension) and the exponential fit for the dynamical gluon mass.}
\label{fig3} 
\end{figure}

It is not difficult to imagine how the csb will depend on the variation
of $K_{F}$ with the number of flavors. The string tension in the
confining propagator plays the same role of the coupling constant
in the one-gluon exchange gap equation, and for $K_{F}$ below a certain
critical value csb will be restored. In the Fig.(\ref{fig3}) we show
the behavior of the dynamical quark masses as a function of the momenta
for different number of flavors, computed with the Gaussian fit for
the string tension and the exponential fit for the dynamical gluon
mass. As the string tension value decreases we have smaller dynamical
masses. 

\begin{figure}[h]
\includegraphics[scale=0.65]{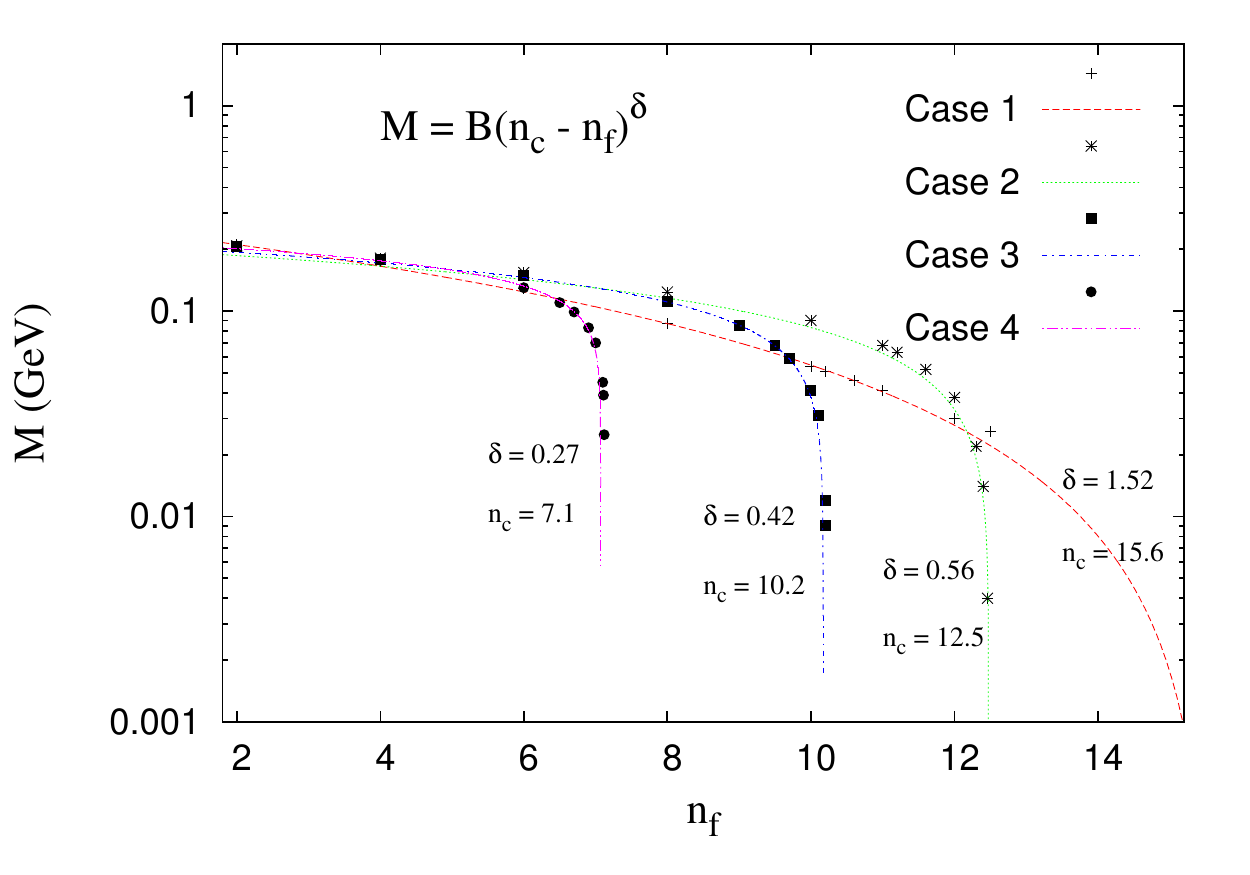} 
\caption{Infrared values of the dynamical quark masses as a function of the
number of quarks. The points corresponding to the central values of
the infrared quark mass were computed with the complete gap equation
considering the different cases of Eq.(7) for the string tension.
The influence of the dynamical gluon mass is negligible compared to
the uncertainty in the string tension value. The fits are given just
to guide the eyes. The results were fitted by the expression $M=A(n_{c}-n_{f})^{\delta}$,
with $n_{c}$ and $\delta$ shown in the figure. Case 2 shows an approximate
mean field behavior near criticality, with a critical number of fermions
$n_{c}\approx12.48$. }
\label{fig4} 
\end{figure}

Fig.(\ref{fig4}) contains the infrared values of the dynamical quark
masses as a function of the number of quarks. These curves were computed
with the complete gap equation considering the fits proposed for the
string tension combined with the exponential fit for the dynamical
gluon mass. We may verify that the chiral symmetry is restored for
$n_{f}$ values of the order of $7$ to $13$ where the uncertainty
is basically due to our poor knowledge of the string tension and the
gluon mass at large $n_{f}$. Note that $m_{g}$ values are not so
relevant for the determination of the critical $n_{f}$ value, as
could be expected in face of the results presented in Refs. \cite{us1,us3}.
This is the main reason why we only focused on one of the fits for
$m_{g}$.

The dynamical masses as a function of the number of fermions shown
in Fig.(4) were fitted with the expression 

\[ M=A(n_{c}-n_{f})^{\delta}, \]

where $n_{c}$ is the critical number of fermions. We can note that
the Case 2 (linear fit for $K_{F}$) leads to a critical number of
fermion $n_{c}\approx12.48$, which is in agreement with lattice data
\cite{for,tom}. In this particular case the critical behavior approaches
a mean field one. Case 1 leads to a larger critical number, but this
was expected because this is the case of a Gaussian fit for the string
tension, and away from the peak the Gaussian tail falls slowly, and
probably this is a bad fit for the string tension at large $n_{f}$
although this is a possible fit for the data. Unfortunately, it is
impressive that there are not lattice data (and with small errors)
for $K_{F}$ at large $n_{f}$. This lack of data does not allow us
to perform more precise calculations. Note that we should not
expect that the Gaussian tail would give a good description of the
critical $K_F$ behavior at large $n_f$.

It has been argued by Maris \cite{Maris} that fermion de-confinement
can be studied through the analytic properties of the fermion propagator
and, in particular, by its Schwinger function 

\[
\Omega(t)=\int d^{3}x\int\frac{d^{4}p}{(2\pi)^{4}}e^{p^{0}t+\vec{p}\cdot\vec{x}}\frac{M(p^{2})}{p^{2}+M^{2}(p^{2})}.
\]

If there are two complex conjugate singularities the Schwinger function
will show an oscillatory behavior, and if it exhibits a real mass singularity
of the propagator, the quark is observable and consequently not confined.
In this case, if there is a stable asymptotic state associated with
the quark propagator with mass $m$ we shall have 

\begin{equation}
\Omega(t)\propto e^{-mt}\rightarrow\lim_{t\rightarrow\infty}\ln\Omega(t)\sim-mt.
\label{ee}
\end{equation}

In Fig.(5) we plot the behavior $\log{\Omega(t)}$ as a function of
$t$ for the case 2 mentioned before. We expect that approaching the critical fermion number the oscillatory behavior of $\Omega(t)$ will never takes place and quarks will be
deconfined corresponding to stable asymptotic states. Another confinement-de-confinement order parameter, as suggested in Ref.\cite{bashir}, is obtained from

\be 
\nu (n_f ) = \frac{1}{\tau (n_f)},
\label{nucon}
\ee

where $\tau (n_f)$ is the location of the first zero of the Schwinger function and its vanishing signals de-confinement, as shown in Fig.(\ref{fig6}).

\begin{figure}[h]
\includegraphics[scale=0.65]{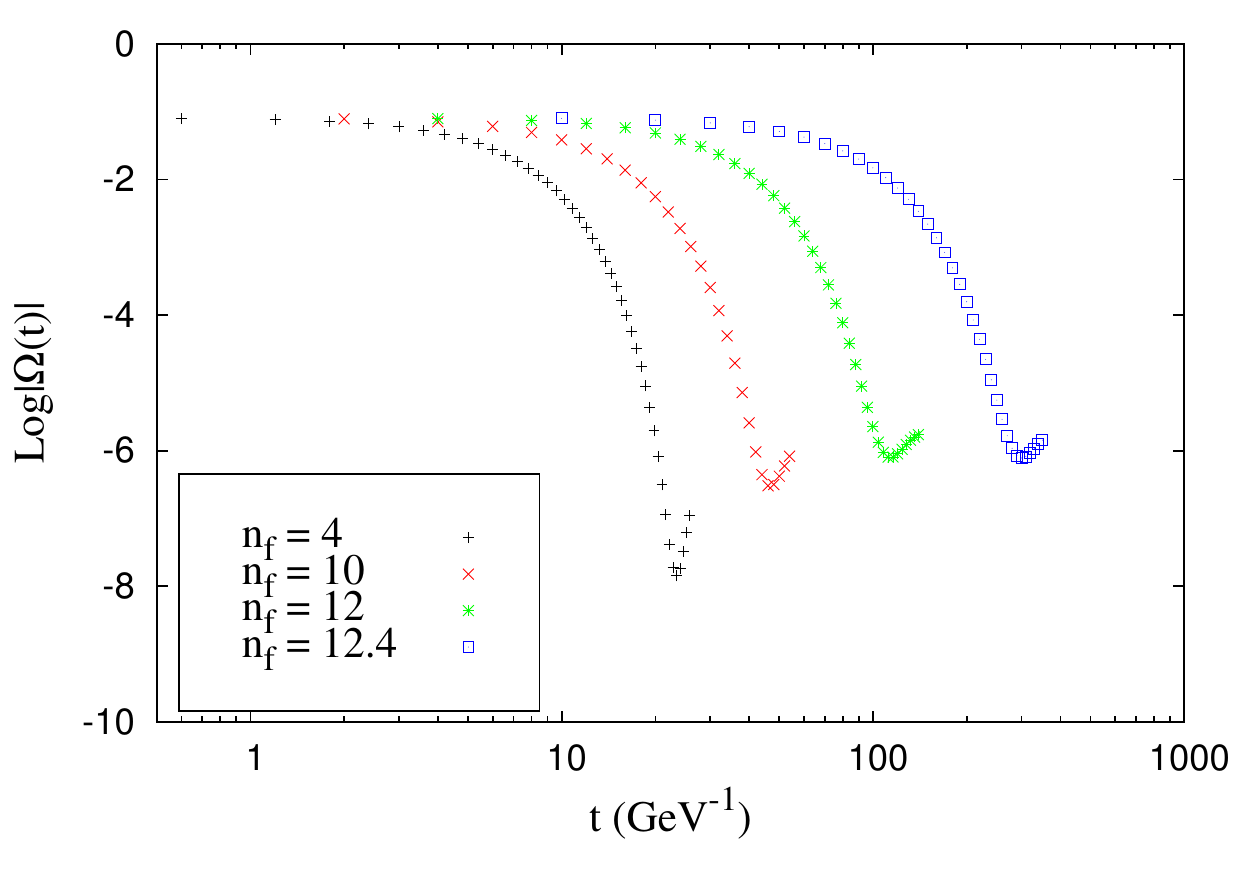} 
\caption{Logarithm of the absolute value of the Schwinger function for the case 2. Note that as the number of fermions is increased the behavior
is pushed to larger $t$ values becoming more and more spaced, and
we expect that approaching the critical number the oscillatory behavior
will never takes place and quarks will be deconfined corresponding
to stable asymptotic states.}
\label{fig5} 
\end{figure}

\begin{figure}[h]
\includegraphics[scale=0.65]{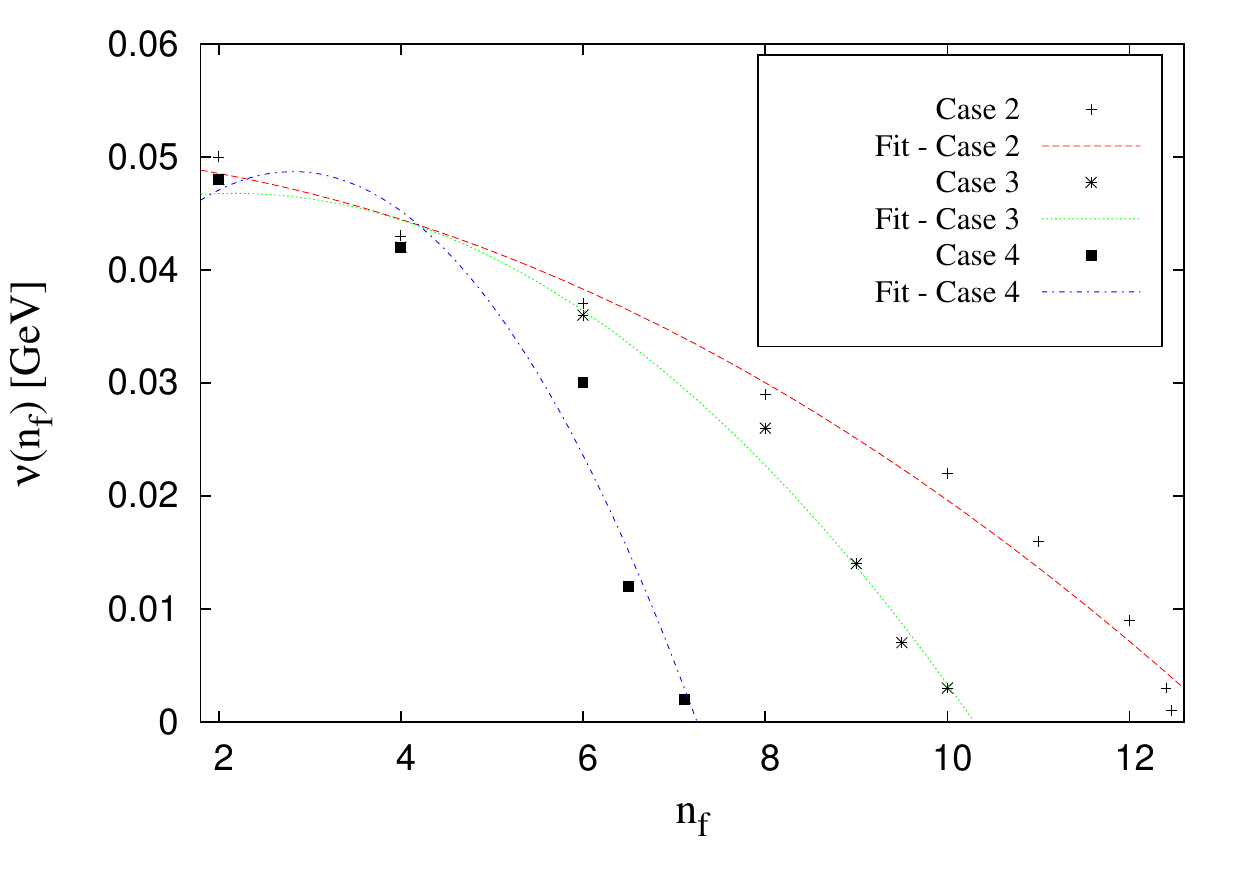} 
\caption{Behavior of the confinement-de-confinement parameter $\nu (n_f ) = {1}/{\tau (n_f)}$ \cite{bashir}, as
a function of the number of flavors in the main cases discussed in the text. Its vanishing indicates de-confinement.}
\label{fig6} 
\end{figure}

In the context of a csb model where the gap equation contains an effective
confining propagator and a dressed gluon propagator with a dynamically
generated mass, we verified that the chiral symmetry is restored for
a large number of quarks. We first discussed the properties of the
model, indicating that the introduction of the effective confining
propagator is one way to introduce confinement by vortices, which
cannot appear in the SDE when we consider only the exchange of dynamically
massive gluons \cite{co2,co3}. We also discussed possible differences
between this and other models to explain csb with dynamically massive
gluons. Our results indicate that the chiral symmetry is restored
for $n_{f}\approx7-13$, in agreement with lattice results \cite{for,tom}.
The values of the string tension ($K_{F}$) and dynamical gluon masses
($m_{g}$) were extrapolated from small to large number of quarks.
This extrapolation is the larger source of uncertainty in our calculation.
The gap equation calculation is quite numerically sensitive to the
decrease of parameters that define the critical bifurcation behavior
(factors like $K_{F}$ or $C_{2}\bar{g}^{2}$ in the one-gluon gap
equation), but certainly the $K_{F}$ extrapolated values is at the
origin of the range of $n_{f}$ critical values. It is important to
have new QCD simulations of the string tension and dynamical gluon
masses for a large number of flavors and for different fermionic representations,
otherwise it will not be possible to perform precise gap equation
calculations of the csb phase diagram as a function of $n_{f}$.

\section*{Acknowledgments}

We thank an anonymous referee for the comments that helped to improve the
manuscript. This research was partially supported by the University of Notre Dame,
by the Conselho Nac. de Desenv. Científico e Tecnológico (CNPq), by
grant 2013/22079-8 of Fundação de Amparo à Pesquisa do Estado de São
Paulo (FA\-PES\-P) and by Coordenação de Aper\-fei\-çoa\-mento
de Pessoal de Nível Superior (CAPES).


\section*{References}

\begin{thebibliography}{10}

\bibitem{lat1} H. Reinhardt, O. Schröder, T. Tok and V. C. Zhukovsky, Phys. Rev. D \textbf{66}, 085004 (2002).

\bibitem{lat2} J. Gattnar, C. Gattringer, K. Langfeld, H. Reinhardt,
A. Schafer, S. Solbrig and T. Tok, Nucl. Phys. B \textbf{716}, 105
(2005).

\bibitem{lat3} P. de Forcrand and M. D'Elia, Phys. Rev. Lett. \textbf{82},
4582 (1999); P. O. Bowman \textit{et al.}, Phys. Rev. D \textbf{78},
054509 (2008).

\bibitem{fa} M. Faber and R. Hollvieser, Acta Phys. Polon. Supp.
\textbf{7}, 457 (2014); R. Hollwieser, M. Faber, T. Schweigler and
U. M. Heller, PoS Lattice \textbf{2013}, 505 (2013); R. Hollwieser,
T. Schweigler, M. Faber and U. M. Heller, Phys. Rev. D \textbf{88},
114505 (2013).

\bibitem{ca} A. Casher, Phys. Lett. B \textbf{83}, 395 (1979).

\bibitem{ba} T. Banks and A. Casher, Nucl. Phys. B \textbf{169}, 103 (1980).

\bibitem{co} J. M. Cornwall, Phys. Rev. D \textbf{22}, 1452 (1980).

\bibitem{su} H. Suganuma, T. M. Doi and T. Iritani, arXiv:1405.1289;
1404.6494; PoS (QCD-TNT-III) 042 (2014); S. Gongyo, T. Iritani and
H. Suganuma, Phys. Rev. D \textbf{86}, 034510 (2012).

\bibitem{co1} J. M. Cornwall, \textit{Center vortices, the functional
Schr}ö\textit{dinger equation, and CSB}, Invited talk at the conference
``Approaches to Quantum Chromodynamics'', Oberwölz, Austria, September
2008, arXiv:0812.0359 {[}hep-ph{]}.

\bibitem{co2} J. M. Cornwall, Phys. Rev. D \textbf{83}, 076001 (2011)

\bibitem{us1} A. Doff, F. A. Machado and A. A. Natale, Annals Phys. \textbf{327}, 1030 (2012).

\bibitem{us2} A. Doff, F. A. Machado and A. A. Natale, New J. Phys. \textbf{14}, 103043 (2012).

\bibitem{us3} R. M. Capdevilla, A. Doff and A. A. Natale, Phys. Lett. B \textbf{728}, 626 (2014).

\bibitem{nat} A. C. Aguilar, A. Mihara and A. A. Natale, Phys. Rev. D \textbf{65}, 054011 (2002).

\bibitem{ha} B. Haeri and M. B. Haeri, Phys. Rev. D \textbf{43}, 3732 (1991).

\bibitem{hab} A. A. Natale and P. S. Rodrigues da Silva, Phys. Lett. B \textbf{392}, 444 (1997); Phys. Lett. B \textbf{390}, 378 (1997).

\bibitem{co3} J. M. Cornwall, Phys. Rev. D \textbf{26}, 1453 (1982).

\bibitem{papa} A. C. Aguilar, D. Binosi and J. Papavassiliou, Phys.
Rev. D \textbf{89}, 085032 (2014); A. C. Aguilar, D. Binosi, D. Iba${\rm \tilde{n}}$ez
and J. Papavassiliou, Phys. Rev. D \textbf{89}, 085008 (2014); A.
C. Aguilar, D. Binosi and J. Papavassiliou, JHEP \textbf{1201}, 050
(2012); A. C. Aguilar, D. Binosi and J. Papavassiliou, Phys. Rev.
D \textbf{84}, 085026 (2011); A. C. Aguilar and J. Papavassiliou,
Phys. Rev. D \textbf{81}, 034003 (2010); A. C. Aguilar, D. Binosi,
J. Papavassiliou and J. Rodriguez-Quintero, Phys. Rev. D \textbf{80},
085018 (2009); A. C. Aguilar, D. Binosi and J. Papavassiliou, Phys.
Rev. D \textbf{78}, 025010 (2008); A. C. Aguilar and J. Papavassiliou,
JHEP \textbf{0612}, 012(2006).

\bibitem{papa1} D. Binosi and J. Papavassiliou, Phys. Rept. \textbf{479}, 1 (2009).

\bibitem{papa2} J. M. Cornwall, J. Papavassiliou and D. Binosi, ``The
Pinch Technique and its Applications to Non-Abelian Gauge Theories'',
Cambridge University Press, 2011.

\bibitem{cuc} A. Cucchieri and T. Mendes, PoS QCD-TNT \textbf{09},
031 (2009); Phys. Rev. Lett. \textbf{100}, 241601 (2008); Phys. Rev.
D \textbf{81}, 016005 (2010); I. Bogolubsky, E. Ilgenfritz, M. Muller-Preussker
and A. Sternbeck, Phys. Lett. B \textbf{676}, 69 (2009).

\bibitem{papa3} A. C. Aguilar and J. Papavassiliou, Phys. Rev. D \textbf{83}, 014013 (2011).

\bibitem{cornwall4} J. M. Cornwall, Mod. Phys. Lett. A \textbf{27}, 1230011 (2012).

\bibitem{bali} G. S. Bali et al. {[}SESAM Collaboration{]}, Phys. Rev. D \textbf{71}, 114513 (2005).

\bibitem{aguina} A. C. Aguilar and A. A. Natale, JHEP \textbf{0408}, 057 (2004).

\bibitem{bazavov} A. Bazavov et al., Phys. Rev. D \textbf{80}, 014504 (2009).

\bibitem{aoki} Y. Aoki et al., JHEP \textbf{0906}, 088 (2009).

\bibitem{karsch} F. Karsch and M. Lutgemeier, Nucl. Phys. B \textbf{550}, 449 (1999).

\bibitem{engels} J. Engels, S. Holtmann and T. Schulze, Nucl. Phys. B \textbf{724}, 357 (2005).

\bibitem{bilgici} E. Bilgici, C. Gattringer, E.-M. Ilgenfritz and A. Maas, JHEP \textbf{0911}, 035 (2009).

\bibitem{for} Ph. de Forcrand, S. Kim and W. Unger, JHEP \textbf{1302}, 051 (2013).

\bibitem{tom} E. T. Tomboulis, Phys. Rev. D \textbf{87}, 034513 (2013).

\bibitem{bas} A. Ayala, A. Bashir, D. Binosi, M. Cristoforetti, and
J. Rodriguez-Quintero, Phys. Rev. D \textbf{86}, 074512 (2012).

\bibitem{kar} F. Karsch and E. Laermann, Rep. Prog. Phys. \textbf{56},
1347 (1993); F. Karsch, E. Laermann and A. Peikert, Nucl. Phys. B
\textbf{605}, 579 (2001); F. Karsch, Lect. Notes Phys. \textbf{583},
209 (2002).

\bibitem{luna} E. G. S. Luna, Phys. Lett. B \textbf{641}, 171 (2006);
E. G. S. Luna and A. A. Natale, Phys. Rev. D \textbf{73}, 074019 (2006);
A. A. Natale, PoS QCD-TNT \textbf{09}, 031 (2009); F. Halzen, G. I.
Krein and A. A. Natale, Phys. Rev. D \textbf{47}, 295 (1993).

\bibitem{Maris} P. Maris, Phys. Rev. D \textbf{52}, 6087\textendash 6097 (1995).

\bibitem{bashir} A. Bashir, A. Raya, S. S\'anchez-Madrigal and C. D. Roberts, Few-Body Syst. \textbf{46}, 229 (2009). 

\end{thebibliography}
\end{document}